\documentclass[10pt,conference]{IEEEtran}

\IEEEoverridecommandlockouts

\usepackage[utf8]{inputenc}
\usepackage{amsmath,amssymb,amsfonts,mathtools,amsthm}
\usepackage{algorithmic}
\usepackage{listings}
\usepackage{graphicx}
\usepackage{textcomp}
\usepackage{xcolor,color,colortbl}
\usepackage{xspace}
\usepackage{hyperref}
\usepackage{lipsum}
\usepackage{multirow,multicol}
\usepackage{tabularx}
\usepackage{rotating}
\usepackage{pifont}
\usepackage{lscape}
\usepackage{adjustbox}
\usepackage{booktabs}
\usepackage{wrapfig}
\usepackage{cleveref}
\usepackage{setspace}
\usepackage{tcolorbox}
\usepackage{url}
\usepackage[font=footnotesize]{caption}

\def\tablename{Table}
\renewcommand{\thetable}{\arabic{table}}

\def\BibTeX{{\rm B\kern-.05em{\sc i\kern-.025em b}\kern-.08em
    T\kern-.1667em\lower.7ex\hbox{E}\kern-.125emX}}

\begin{document}

\crefformat{section}{\S#2#1#3}
\crefformat{subsection}{\S#2#1#3}
\crefformat{subsubsection}{\S#2#1#3}

\newtheorem{definition}{Definition}

% mutators
\newcommand{\mathWeight}{{\small \textsf{MATH\_WEIGHT}}\xspace}
\newcommand{\mathWeightConv}{{\small \textsf{MATH\_WEIGHT\_CONV}}\xspace}
\newcommand{\mathActWeight}{{\small \textsf{MATH\_ACT\_WEIGHT}}\xspace}
\newcommand{\mathLSTMInWeight}{{\small \textsf{MATH\_LSTM\_IN\_WEIGHT}}\xspace}
\newcommand{\mathLSTMForgetWeight}{{\small \textsf{MATH\_LSTM\_FORGET\_WEIGHT}}\xspace}
\newcommand{\mathLSTMCellWeight}{{\small \textsf{MATH\_LSTM\_CELL\_WEIGHT}}\xspace}
\newcommand{\mathLSTMOutWeight}{{\small \textsf{MATH\_LSTM\_OUT\_WEIGHT}}\xspace}
\newcommand{\mathBias}{{\small \textsf{MATH\_BIAS}}\xspace}
\newcommand{\deleteLayer}{{\small \textsf{DEL\_LAYER}}\xspace}
\newcommand{\duplicateLayer}{{\small \textsf{DUP\_LAYER}}\xspace}
\newcommand{\mathConvBias}{{\small \textsf{MATH\_CONV\_BIAS}}\xspace}
\newcommand{\mathLSTMInBias}{{\small \textsf{MATH\_LSTM\_IN\_BIAS}}\xspace}
\newcommand{\mathLSTMForgetBias}{{\small \textsf{MATH\_LSTM\_FORGET\_BIAS}}\xspace}
\newcommand{\mathLSTMCellBias}{{\small \textsf{MATH\_LSTM\_CELL\_BIAS}}\xspace}
\newcommand{\mathLSTMOutBias}{{\small \textsf{MATH\_LSTM\_OUT\_BIAS}}\xspace}
\newcommand{\actRep}{{\small \textsf{ACT\_FUNC\_REP}}\xspace}
\newcommand{\mathPoolSz}{{\small \textsf{MATH\_POOL\_SZ}}\xspace}
\newcommand{\mathStrides}{{\small \textsf{MATH\_STRIDES}}\xspace}
\newcommand{\mathKernelSz}{{\small \textsf{MATH\_KERNEL\_SZ}}\xspace}
\newcommand{\mathFilters}{{\small \textsf{MATH\_FILTERS}}\xspace}
\newcommand{\paddingRep}{{\small \textsf{PADDING\_REP}}\xspace}
\newcommand{\recActRep}{{\small \textsf{REC\_ACT\_FUNC\_REP}}\xspace}

% acronyms
\newcommand{\tool}{{\footnotesize \textsf{deepmufl}}\xspace}
\newcommand{\toolSmall}{{\scriptsize \textsf{deepmufl}}\xspace}
\newcommand{\toolFNS}{{\footnotesize \textsf{deepmufl}}\xspace}

% techniques
\newcommand{\dd}{DeepDiagnosis\xspace}
\newcommand{\dl}{DeepLocalize\xspace}
\newcommand{\umlaut}{UMLAUT\xspace}
\newcommand{\deepfd}{DeepFD\xspace}
\newcommand{\neuralint}{Neuralint\xspace}
\newcommand{\metallaxis}{Metallaxis\xspace}
\newcommand{\muse}{MUSE\xspace}
\newcommand{\autotrainer}{AutoTrainer\xspace}

% numbers
\newcommand{\allMutators}{79\xspace}
\newcommand{\bugsCount}{109\xspace}
\newcommand{\toolDetected}{53\xspace}
\newcommand{\toolAvgTime}{1,040.49\xspace}
\newcommand{\toolAvgTimeHalf}{562.72\xspace}
\newcommand{\toolFoundBugsHalf}{49\xspace}
\newcommand{\soPosts}{8,412\xspace}

% foreign
\newcommand{\ie}{\textit{i.e.}\xspace}
\newcommand{\eg}{\textit{e.g.}\xspace}
\newcommand{\etc}{\textit{etc}\xspace}
\newcommand{\perse}{\textit{per se}\xspace}
\newcommand{\ala}{\textit{à la}\xspace}
\newcommand{\cf}{\textit{c.f.}\xspace}
\newcommand{\via}{\textit{via}\xspace}
\newcommand{\vs}{\textit{vs.}\xspace}
\newcommand{\etal}{\textit{et al.}\xspace}
\newcommand{\viceversa}{\textit{vice versa}\xspace}

% symbols
\newcommand{\cmark}{\ding{51}\xspace}
\newcommand{\xmark}{\ding{55}\xspace}

% colors
\definecolor{GrayOne}{gray}{0.9}
\definecolor{GrayTwo}{gray}{0.8}
\definecolor{GrayThree}{gray}{0.7}
\definecolor{GrayFour}{gray}{0.6}
\definecolor{GrayFive}{gray}{0.5}
\definecolor{codegreen}{rgb}{0,0.6,0}
\definecolor{codegray}{rgb}{0.5,0.5,0.5}
\definecolor{codepurple}{rgb}{0.58,0,0.82}
\definecolor{backcolour}{rgb}{0.95,0.95,0.92}
\definecolor{GreenClover}{rgb}{0.71,0.88,0.72}
\definecolor{MistyRose}{rgb}{1.0,0.89,0.88}

% pens
\newcommand{\ali}[1]{\textcolor[rgb]{0.0,0.0,1.0}{#1}}
\newcommand{\hridesh}[1]{\textcolor[rgb]{1.0,0.0,0.0}{#1}}
\newcommand{\arbab}[1]{\textcolor[rgb]{0.0,1.0,0.0}{#1}}

% python lst

\lstdefinestyle{mystyle}{
	backgroundcolor=\color{backcolour},   
	commentstyle=\color{codegreen},
	keywordstyle=\color{magenta},
	numberstyle=\tiny\color{codegray},
	stringstyle=\color{codepurple},
	basicstyle=\scriptsize,
	breakatwhitespace=false,         
	breaklines=true,                 
	captionpos=b,                    
	keepspaces=true,                 
	numbers=left,                    
	numbersep=5pt,                  
	showspaces=false,                
	showstringspaces=false,
	showtabs=false,                  
	tabsize=2
}
\lstset{style=mystyle}

% symbols
\newcommand{\nonviable}{\ensuremath{\text{\ding{109}}}\xspace}%
\newcommand{\impacts}{\ensuremath{\text{\ding{51}}}\xspace}%
\newcommand{\noimpact}{\ensuremath{\text{\ding{55}}}\xspace}%

\title{Mutation-based Fault Localization\\of Deep Neural Networks}

% \author{\IEEEauthorblockN{Anonymous Author(s)}}
\author{\IEEEauthorblockN{Ali Ghanbari}
\IEEEauthorblockA{
\textit{Dept. of Computer Science}\\
\textit{Iowa State University}\\
Ames, Iowa, USA\\
\url{alig@iastate.edu}
}

\and

\IEEEauthorblockN{Deepak-George Thomas}
\IEEEauthorblockA{
\textit{Dept. of Computer Science}\\
\textit{Iowa State University}\\
Ames, Iowa, USA\\
\url{dgthomas@iastate.edu}
}

\and

\IEEEauthorblockN{Muhammad Arbab Arshad}
\IEEEauthorblockA{
\textit{Dept. of Computer Science}\\
\textit{Iowa State University}\\
Ames, Iowa, USA\\
\url{arbab@iastate.edu}
}

\and

\IEEEauthorblockN{Hridesh Rajan}
\IEEEauthorblockA{
\textit{Dept. of Computer Science}\\
\textit{Iowa State University}\\
Ames, Iowa, USA\\
\url{hridesh@iastate.edu}
}

}

\thispagestyle{plain}
\pagestyle{plain}

\maketitle

\begin{abstract}
Deep neural networks (DNNs) are susceptible to bugs, just like other types of software systems.
A significant uptick in using DNN, and its applications in wide-ranging areas, including safety-critical systems, warrant extensive research on software engineering tools for improving the reliability of DNN-based systems.
One such tool that has gained significant attention in the recent years is \textit{DNN fault localization}.
This paper revisits \textit{mutation-based fault localization} in the context of DNN models and proposes a novel technique, named \toolFNS, applicable to a wide range of DNN models.
We have implemented \toolFNS and have evaluated its effectiveness using \bugsCount bugs obtained from StackOverflow.
Our results show that \toolFNS detects \toolDetected/\bugsCount of the bugs by ranking the buggy layer in top-1 position, outperforming state-of-the-art static and dynamic DNN fault localization systems that are also designed to target the class of bugs supported by \toolFNS.
Moreover, we observed that we can halve the fault localization time for a pre-trained model using \textit{mutation selection}, yet losing only 7.55\% of the bugs localized in top-1 position.
\end{abstract}


\section{Introduction}\label{sec:introduction}
Software bugs~\cite{ieeestd} are a common and costly problem in modern software systems, costing the global economy billions of dollars annually~\cite{mcpeak2017bugs}.
Recently, data-driven solutions have gained significant attention for their ability to efficiently and cost-effectively solve complex problems.
With the advent of powerful computing hardware and an abundance of data, the use of deep learning~\cite{lecun2015deep}, which is based on deep neural networks (DNNs), has become practical.
Despite their increasing popularity and success stories, DNN models, like any other software, may contain bugs~\cite{islam2019comprehensive,islam2020repairing,zhang2018empirical,humbatova2020taxonomy}, which can undermine their safety and reliability in various applications.
Detecting DNN bugs is \textit{not} easier than detecting bugs in traditional programs, \ie, programs without any data-driven component in them, as DNNs depend on the properties of the training data and numerous hyperparameters~\cite{wardat2021deepdiagnosis}.
Mitigating DNN bugs has been the subject of fervent research in recent years, and various techniques have been proposed for testing~\cite{pei2019deepxplore,kim2019guiding}, fault localization~\cite{wardat2021deeplocalize,nikanjam2021automatic}, and repair~\cite{usman2021nn,zhang2021autotrainer} of DNN models.
% Software bugs~\cite{ieeestd} are common in modern software systems, costing the global economy billions of dollars every year~\cite{mcpeak2017bugs}.
% Data-driven software solutions~\cite{biswas2022art} have lately gained attention from academia and industry due to their capabilities in solving complex problems efficiently and cheaper.
% With the advent of powerful computing hardware, and abundance of data, the use of deep learning~\cite{lecun2015deep}, which is based on deep neural networks (DNN), has become practical.
% Despite the increasing popularity and remarkable success stories, DNN models, like any other software, may have bugs~\cite{islam2019comprehensive,islam2020repairing,zhang2018empirical,humbatova2020taxonomy}, thereby undermining its safety and reliability in many applications.
% Detecting DNN bugs is \textit{not} easier than detecting bugs in classic programs, \ie, programs without any data-driven component, as DNN programs depend on the properties of training data and tens of hyperparameters~\cite{wardat2021deepdiagnosis}.
% Mitigating DNN bugs has been the topic of intense research in the past decade and many researchers have proposed techniques for testing~\cite{pei2019deepxplore,kim2019guiding}, fault localization~\cite{wardat2021deeplocalize,nikanjam2021automatic}, and repair~\cite{usman2021nn,zhang2021autotrainer} of DNN models.

Fault localization in the context of traditional programs has been extensively studied~\cite{wong2016survey}, with one well-known approach being \textit{mutation-based fault localization} (MBFL)~\cite{papadakis2012using,moon2014ask}.
This approach is based on mutation analysis~\cite{demillo1978hints}, which is mainly used to assess the quality of a test suite by measuring the ratio of artificially introduced bugs that it can detect.
MBFL improves upon the more traditional, lightweight \textit{spectrum-based fault localization}~\cite{abreu2007accuracy,dallmeier2005lightweight,jones2002visualization,naish2011model,yoo2012evolving,xie2013provably} by uniquely capturing the relationship between individual statements in the program and the observed failures.
While both spectrum-based fault localization~\cite{ma2018mode,eniser2019deepfault} and mutation analysis~\cite{humbatova2021deepcrime,hu2019deepmutation++,ma2018deepmutation} have been studied in the context of DNNs, to the best of our knowledge, MBFL for DNNs has not been explored by the research community, yet the existing MBFL approaches are not directly applicable to DNN models.
% In fault localization of classic programs, mutation-based fault localization~\cite{papadakis2012using,moon2014ask} (MBFL) is based on \textit{mutation analysis}~\cite{demillo1978hints}, a software analysis approach to assess the quality of the test suite by measuring the ratio of detectable artificial bugs injected in the program.
% MBFL improves upon the more traditional, lightweight spectrum-based fault localization~\cite{abreu2007accuracy,dallmeier2005lightweight,jones2002visualization,naish2011model,yoo2012evolving,xie2013provably}, by uniquely capturing the relationship between individual statements in the program and the observed failure.
% In the context of DNN, spectrum-based fault localization~\cite{ma2018mode,eniser2019deepfault}, as well as mutation analysis~\cite{humbatova2021deepcrime,hu2019deepmutation++,ma2018deepmutation}, has been studied.
% However, to the best of our knowledge, MBFL for DNN has not received any attention from the research community, so far, while the existing MBFL approaches are not directly applicable to DNN models.

This paper revisits the idea of MBFL in the context of DNNs.
Specifically, we design, implement, and evaluate a technique, named \tool, to conduct MBFL in pre-trained DNN models.
The basic idea behind \tool is derived from its traditional MBFL counterparts, namely, \metallaxis~\cite{papadakis2015metallaxis} and \muse~\cite{moon2014ask}, that are based on measuring the impact of mutations on passing and failing test cases (see~\cref{sec:background} for more details).
In summary, given a pre-trained model and a set of data points, \tool separates the data points into two sets of ``passing'' and ``failing'' data points (test cases), depending on whether the output of the model matches the ground-truth.
\tool then localizes the bug in two phases, namely \textit{mutation generation phase} and \textit{mutation testing/execution phase}.
In mutation generation phase, it uses \allMutators mutators, a.k.a. mutation operators, to systematically mutate the model, \eg, by replacing activation function of a layer, so as to generate a pool of mutants, \ie, model variants with seeded bugs.
In mutation testing phase, \tool feeds each of the mutants with passing and failing data points and compares the output to the output of the original model to record the number of passing and failing test cases that are impacted by the injected bugs.
In this paper, we study two types of impacts: \textit{type 1} impact, \ala \muse, which tracks only fail to pass and pass to fail, and \textit{type 2} impact, like \metallaxis, which tracks changes in the actual output values.
\tool uses these numbers to calculate \textit{suspiciousness values} for each layer according to \muse, as well as two variants of \metallaxis formulas.
The layers are then sorted in descending order of their suspiciousness values for the developer to inspect.

We have implemented \tool on top of Keras~\cite{chollet2015keras}, and it supports three types of DNN models for regression, as well as classification tasks that must be written using {\small \texttt{Sequential}} API of Keras: fully-connected DNN, convolutional neural network (CNN), and recurrent neural network (RNN).
Extending \tool to other libraries, \eg, TensorFlow~\cite{tensorflow2016abadi} and PyTorch~\cite{pytorch}, as well as potentially to other model architectures, \eg, functional model architecture in Keras, is a matter of investing engineering effort on the development of new mutators tailored to such libraries and models.
Since the current implementation of \tool operates on pre-trained models, its scope is limited to \textit{model bugs}~\cite{humbatova2020taxonomy}, \ie, bugs related to activation function, layer properties, model properties, and bugs due to missing/redundant/wrong layers (see~\cref{sec:evaluation}).
%for more details).
% Extending \tool to perform mutation generation and execution at the source code level is possible and is left as a future work.

We have evaluated \tool using a diverse set of \bugsCount Keras bugs obtained from StackOverflow.
These bugs are representatives of the above-mentioned model bugs, in that our dataset contains examples of each bug sub-category at different layers of the models suited for different tasks.
For example, concerning the sub-category \textit{wrong activation function} model bug, we have bugs in regression and classification fully-connected DNN, CNN, and RNN models that have wrong activation function of different types (\eg, ReLU, softmax, \etc.) at different layers.
For \toolDetected of the bugs, \tool, using its \muse configuration, pinpoints the buggy layer by ranking it in top-1 position.
We have compared \tool's effectiveness to that of state-of-the-art static and dynamic DNN fault localization systems \neuralint~\cite{nikanjam2021automatic}, \dl~\cite{wardat2021deeplocalize}, \dd~\cite{wardat2021deepdiagnosis}, and \umlaut~\cite{schoop2021umlaut} that are also designed to detect model bugs.
Our results show that, in our bug dataset, \tool, in its \muse configuration, is 77\% more effective than \dd, which detects 30 of the bugs.
% The fact that \tool does not intercept the training process does \textit{not} make the technique slower than other dynamic systems such as \dl, \dd, \umlaut, or \deepfd.
% Specifically, \tool takes, on average, \toolAvgTime seconds to finish, while in some cases, \dl fails to finish fault localization within our 5-hour time budget.
% This is because \dl incurs heavy overhead due to the instrumentation that it performs.

Despite this advantage of \tool in terms of effectiveness, since it operates on a pre-trained model, it is slower than state-of-the-art DNN fault localization tools from an end-user's perspective.
However, this is mitigated, to some extent, by the fact that similar to traditional programs, one can perform \textit{mutation selection}~\cite{wong1995reducing} to curtail the mutation testing time: we observed that by randomly selecting 50\% of the mutants for testing, we can still find \toolFoundBugsHalf of the bugs in top-1 position, yet we halve the fault localization time after training the model.

In summary, this paper makes the following contributions.
\begin{itemize}
    \item \textbf{Technique}: We develop MBFL for DNN and implement it in a novel tool, named \tool, that can be uniformly applied to a wide range of DNN model types.
    \item \textbf{Study}: We compare \tool to state-of-the-art static and dynamic fault localization approaches and observed:
    \begin{itemize}
        \item In four configurations, \tool outperforms other approaches in terms of the number of bugs that appear in top-1 position and it detects 21 bugs that none of the studied techniques were able to detect.
        \item We can halve the fault localization time for a pre-trained model by random mutation selection without significant loss of effectiveness.
        % \item In our dataset of bugs, \tool, in its \muse configuration, is 3+X more effective than other techniques in reducing the expected user efforts in locating the bug.
    \end{itemize}
    \item \textbf{Bug Dataset}: We have created the largest curated dataset of \textit{model bugs}, comprising 109 Keras models ranging from regression to classification and fully-connected DNN to CNN and RNN.
    %We sieved through \soPosts StackOverflow posts to create the largest curated dataset of \textit{model bugs} for Keras models ranging from regression to classification and fully-connected DNN to CNN and RNN.
\end{itemize}

\noindent
\textbf{Paper organization.}
In the next section, we review concepts of DNNs, mutation analysis, and MBFL.
In~\cref{sec:motivating}, we present a motivating example and discusses how \tool works under the hood.
In~\cref{sec:approach}, we present technical details of the proposed approach, before discussing the scope of \tool in~\cref{sec:scope}.
In~\cref{sec:evaluation}, we present the results of our experiments with \tool and state-of-the-art DNN fault localization tools from different aspects.
We discuss threats to validity in~\cref{sec:threats} and conclude the paper in~\cref{sec:conclusion}.

\noindent
\textbf{Data availability.}
The source code of \tool and the data associated with our experiments are publicly available~\cite{replication}.

\section{Background}\label{sec:background}
\subsection{Mutation Analysis}\label{sec:background:analysis}
Mutation analysis~\cite{demillo1978hints}, is a program analysis method for assessing the quality of the test suite.
It involves generating a pool of program variants, \ie, \textit{mutants}, by systematically mutating program elements, \eg, replacing an arithmetic operator with another, and running the test suite against the mutants to check if the output of the mutated program is different from that of the original one;
if different, the mutant is marked as \textit{killed}, otherwise as \textit{survived}.
A mutant might survive because it is semantically equivalent to the original program, hence the name \textit{equivalent} mutant.
Test suite quality is assessed by computing a \textit{mutation score} for the test suite, which is the ratio of killed mutants over the non-equivalent survived mutants.
Mutation score indicates how good a test suite is in detecting real bugs~\cite{andrews2005mutation}.
%The higher mutation score, the better.
In addition to its original use, mutation analysis has been used for many other purposes~\cite{papadakis2019mutation}, such as fault localization~\cite{papadakis2012using,moon2014ask}, automated program repair~\cite{debroy2010using,ghanbari2019practical}, test generation~\cite{fraser2015achieving,souza2016strong} and prioritization~\cite{shin2019empirical}, program verification~\cite{galeotti2015inferring,groce2015verified}, \etc.

\subsection{Mutation-based Fault Localization}\label{sec:background:localization}
Mutation-based fault localization (MBFL) uses mutation analysis to find bugs.
In this section, we review two major approaches to MBFL, namely \metallaxis~\cite{papadakis2015metallaxis} and \muse~\cite{moon2014ask}.
Both of these approaches are implemented in \tool.
The reader is referred to the original papers \cite{papadakis2015metallaxis,moon2014ask} for examples explicating the rationale behind each approach. 
\subsubsection{\metallaxis}
\metallaxis~\cite{papadakis2015metallaxis} posits that mutants generated by mutating the same program element are likely to exhibit similar behaviors and mutants generated by mutating different program elements are likely to exhibit different behaviors.
Since a fault itself can also be viewed as a mutant, it is expected to behave similar to other mutants generated by mutating that same buggy program element and can be located by examining the mutants based on this heuristic.
\metallaxis assumes that the mutants impacting the test outputs, or their error messages, \eg, stack trace, as \textit{impacting the tests}.
Thus, mutants impacting failing test cases might indicate that their corresponding code elements is the root cause of the test failures, while mutants impacting passing test cases might indicate that their corresponding code elements are correct.
Once the number of impacted passing and failing test cases are calculated, \metallaxis uses a fault localization formula to calculate suspiciousness values for each element.

\metallaxis fault localization formula can be viewed as an extension to that of spectrum-based fault localization, by treating all mutants impacting the tests as covered elements while the others as uncovered elements.
Specifically, the maximum suspiciousness value of the mutants of a corresponding code element is returned as the suspiciousness value of the code element.
More concretely, assuming we are using SBI formula~\cite{liblit2005scalable}, suspiciousness value for a program element $e$, denoted $s(e)$, is calculated as follows.
\begin{equation}\label{eq:metallaxis:sbi}
    s(e) = \max_{m\in M(e)}\left(\frac{|T_f(m,e)|}{|T_f(m,e)| + |T_p(m,e)|}\right),
\end{equation}
where $M(e)$ denotes the set of all mutants targeting program element $e$, $T_f(m,e)$ is the set of failing test cases that are impacted by the mutant $m$, while $T_p(m,e)$ denotes the set of passing test cases that are impacted by $m$.
In this definition, and in the rest of the paper, the notation $|\cdot|$ represents the size of a set.
Alternatively, had we used Ochiai~\cite{abreu2006evaluation}, \metallaxis suspiciousness formula would be modified as follows.
\begin{equation}\label{eq:metallaxis:ochiai}
    s(e) = \max_{m\in M(e)}\left(\frac{|T_f(m,e)|}{\sqrt{(|T_f(m,e)| + |T_p(m,e)|)|T_f|}}\right),
\end{equation}
where
%$M(e)$, $T_f(m,e)$, and $T_p(m,e)$ are defined as before and 
$T_f$ denotes the set of all failing tests cases.

\subsubsection{\muse}
\muse~\cite{moon2014ask} is based on the assumption that mutating a faulty program element is likely to impact more failed test cases than passing test cases by ``fixing'' it, while mutating a correct program element is likely to impact more passing test cases than failing test cases by breaking it.
The notion of ``impacting test cases'' in \muse, unlike \metallaxis, is more rigid, in that it refers to making passing test cases fail, \viceversa.
Once the number of impacted failing and passing test cases are identified, \textit{suspiciousness values} can be calculated using the following formula.
\begin{equation}\label{eq:muse}
    s(e) = \frac{1}{|M(e)|}\Sigma_{m\in M(e)}\left(\frac{|T_f(m,e)|}{|T_f|} - \alpha\frac{|T_p(m,e)|}{|T_p|}\right),
\end{equation}
where
%$M(e)$, $|T_f(m,e)|$, $|T_p(m,e)|$, and $|T_f|$ are defined as before, and
$T_p$ denotes the set of all passing tests cases and $\alpha$ is a constant used to balance the two ratios that is defined to be $\frac{|F\leadsto P|}{|T_f|}\times\frac{|T_p|}{|P\leadsto F|}$.
In the latter definition, $F\leadsto P$ denotes the set of failing test cases that pass due to some mutation, while $P\leadsto F$ denotes the set of passing test cases that fail as a result of some mutation.

\subsection{Deep Neural Networks}
A neural network is intended to compute a function of the form $\mathbb{R}^m\longrightarrow\mathbb{R}^n$, where $m,n$ are positive integers.
A neural network is often represented as a weighted directed acyclic graph arranged in layers of three types, \ie, \textit{input layer}, one or more \textit{hidden layers}, and an \textit{output layer}.
Input and output layers output linear combination of their inputs, while hidden layers can be viewed as more complex computational units, \eg, a \textit{non-linear unit}, \textit{convolutional unit}, or a  \textit{batch normalization unit}.
A non-linear unit is composed of \textit{neurons}, functions applying a non-linear \textit{activation function}, \eg, rectified linear unit (ReLU), tanh, or sigmoid, on the weighted sum of their inputs.
A convolutional layer, calculates the convolution between the vector of the values obtained from the previous layer and a learned kernel matrix.
Lastly, a batch normalization layer, normalizes the vector of values obtained from the previous layer \via centering or re-scaling.
A neural network with two or more hidden layers is referred to as a \textit{deep neural network} (DNN).

\section{Motivating Example}\label{sec:motivating}
In this section, we first describe how \tool helps programmers detect and fix bugs by presenting a hypothetical use case scenario and then motivate the idea behind \tool by describing the details of how \tool works, under the hood, on the example developed in the use case story.

Courtney is a recent college graduate working as a junior software engineer at an oil company, which frequently makes triangular structures, made of epoxy resin, of varying sizes to be used under the water.
The company needs to predict with at least 60\% confidence that a mold of a specific size will result in an epoxy triangle after it has been dried, and potentially shrunk, and it does not need to spend time on cutting and/or sanding the edges.
Over time, through trial and error, the company has collected 1,000 data points of triangle edge lengths and whether or not a mold of that size resulted in a perfect triangle.
Courtney's first task is to write a program that given three positive real numbers $a$, $b$, and $c$, representing the edge lengths of the triangle mold, determines if the mold will result in epoxy edges that form a perfect triangle.
As a first attempt, she writes the program shown in Listing~\ref{lst:first}.
\begin{center}
    \begin{adjustbox}{width=0.8\columnwidth}
\lstset{language=python}
 \begin{lstlisting}[language = python,label=lst:first, basicstyle=\fontsize{6}{9}\selectfont, caption=Courtney's first attempt]
# load 994 of the data points as X_train and y_train
# ...
model = Sequential()
model.add(Dense(2, activation='relu'))
model.add(Dense(2, activation='relu'))

model.compile(loss='sparse_categorical_crossentropy',
    optimizer='adam', metrics=['accuracy'])
model.fit(X_train, y_train, epochs=100, validation_split=0.1)
\end{lstlisting}
\end{adjustbox}
\end{center}

The program uses 994 out of 1,000 data points for training a model.
After testing the model on the remaining 6 data points, she realizes that the model achieves no more than 33\% accuracy.
Fortunately, Courtney uses an IDE equipped with a modern DNN fault localization tool, named \tool, which is known to be effective at localizing bugs that manifest as stuck accuracy/loss.
She uses \tool, with its default settings, \ie, \metallaxis with SBI, to find the faulty part of her program.
The tool receives the fitted model in {\small \texttt{.h5}} format~\cite{h5format} together with a set of testing data points $T$ and returns a ranked list of model elements; layers, in this case.
After Courtney provides \tool with the model saved in \texttt{.h5} format and the 6 testing data points that she had, within a few seconds, the tool returns a list with two items, namely {\small \texttt{Layer 2}} and {\small \texttt{Layer 1}}, corresponding to the lines 5 and 4, respectively, in Listing~\ref{lst:first}.
Once she navigates to the details about {\small \texttt{Layer 2}}, she receives a ranked list with 5 elements, \ie, {\small \texttt{Mutant 12: replaced activation function `relu' with `softmax'}}, ..., {\small \texttt{Mutant 10: divided weights by 2}}, {\small \texttt{Mutant 11: divided bias by 2}}.
By seeing the description of {\small \texttt{Mutant 12}}, Courtney immediately recalls her machine learning class wherein they were advised that in classification tasks they should use \textit{softmax} as the activation function of the last layer.
%\footnote{Note that in binary classification, sigmoid or softmax can be used in the last layer, but in this example softmax should be used, as we have two units in the last layer. We emphasize that this is just a simple example and {\scriptsize \textsf{deepmufl}} tries all activation functions.}.
She then changes the activation function of the last layer at Line 5 of Listing~\ref{lst:first} from {\small \texttt{relu}} to {\small \texttt{softmax}}.
By doing so, the model achieves an accuracy of 67\% on the test dataset, and similarly on a cross-validation, exceeding the expectations of the company.
\begin{table}[t]
  \centering
  \caption{An example of how \toolSmall uses \metallaxis' default formula to localize the bug in the model of Listing~\ref{fig:badDNN}}\label{tab:motivating}%
  \resizebox{\columnwidth}{!}{
    \begin{tabular}{r|r||l||l|l|l|r|r|r||r|r}
    \multicolumn{1}{r|}{\multirow{2}{*}{\textbf{Layer}}} & \multicolumn{1}{r||}{\multirow{2}{*}{\textbf{Neuron}}} & \multicolumn{1}{c||}{\multirow{2}{*}{\textbf{Mutant}}} & \multicolumn{6}{c||}{\textbf{Impact?}}                 & \multicolumn{1}{r|}{\multirow{2}{*}{$\frac{|T_f(m,e)|}{|T_f(m,e)| + |T_p(m,e)|}$}} & \multicolumn{1}{r}{\multirow{2}{*}{\textbf{max}}} \\
\cline{4-9}          &       &       & \multicolumn{1}{c|}{\cellcolor{MistyRose}T1} & \multicolumn{1}{c|}{\cellcolor{MistyRose}T2} & \multicolumn{1}{c|}{\cellcolor{MistyRose}T3} & \multicolumn{1}{c|}{\cellcolor{MistyRose}T4} & \multicolumn{1}{c|}{\cellcolor{GreenClover}T5} & \multicolumn{1}{c||}{\cellcolor{GreenClover}T6} &       &  \\
    \hline
    \multicolumn{1}{c|}{\multirow{6}{*}{L1}} & \multicolumn{1}{c||}{\multirow{3}{*}{N1}} & M1: weights / 2 &       &       &       &       & \multicolumn{1}{l|}{$\bullet$} &       & 0     & \multicolumn{1}{r}{\multirow{6}{*}{0}} \\
\cline{3-10}          &       & M2: bias / 2 &       &       &       &       & \multicolumn{1}{l|}{$\bullet$} &       & 0     &  \\
\cline{3-10}          &       & M3: relu $\rightarrow$ softmax &       &       &       &       & \multicolumn{1}{l|}{$\bullet$} & \multicolumn{1}{l||}{$\bullet$} & 0     &  \\
\cline{2-10}          & \multicolumn{1}{c||}{\multirow{3}{*}{N2}} & M4: weights / 2 &       &       &       &       & \multicolumn{1}{l|}{$\bullet$} &       & 0     &  \\
\cline{3-10}          &       & M5: bias / 2 &       &       &       &       & \multicolumn{1}{l|}{$\bullet$} &       & 0     &  \\
\cline{3-10}          &       & M6: relu $\rightarrow$ softmax &       &       &       &       & \multicolumn{1}{l|}{$\bullet$} & \multicolumn{1}{l||}{$\bullet$} & 0     &  \\
    \hline
    \multicolumn{1}{c|}{\multirow{6}{*}{L2}} & \multicolumn{1}{c||}{\multirow{3}{*}{N3}} & M7: weights / 2 &       &       & $\bullet$     & \multicolumn{1}{l|}{$\bullet$} &       & \multicolumn{1}{l||}{$\bullet$} & 0.67  & \multicolumn{1}{r}{\multirow{6}{*}{1}} \\
\cline{3-10}          &       & M8: bias / 2 &       & $\bullet$   & $\bullet$     &       & \multicolumn{1}{l|}{$\bullet$} &       & 0.67  &  \\
\cline{3-10}          &       & \cellcolor{GrayThree}M9: relu $\rightarrow$ softmax & \cellcolor{GrayThree}$\bullet$ & \cellcolor{GrayThree}$\bullet$ & \cellcolor{GrayThree}$\bullet$ & \cellcolor{GrayThree} & \cellcolor{GrayThree} & \cellcolor{GrayThree} & \cellcolor{GrayThree}1 &  \\
\cline{2-10}          & \multicolumn{1}{c||}{\multirow{3}{*}{N4}} & M10: weights / 2 &       & $\bullet$     & $\bullet$     &       &       & \multicolumn{1}{l||}{$\bullet$} & 0.67  &  \\
\cline{3-10}          &       & M11: bias / 2 &       & $\bullet$     &       &       & \multicolumn{1}{l|}{$\bullet$} &       & 0.5   &  \\
\cline{3-10}          &       & \cellcolor{GrayThree}M12: relu $\rightarrow$ softmax & \cellcolor{GrayThree}$\bullet$ & \cellcolor{GrayThree}$\bullet$ & \cellcolor{GrayThree}$\bullet$ & \cellcolor{GrayThree} & \cellcolor{GrayThree} & \cellcolor{GrayThree} & \cellcolor{GrayThree}1 &  \\
\hline
    \end{tabular}%
    }
\end{table}%

\begin{wrapfigure}{l}{0.17\textwidth}
\centering
    \includegraphics[scale=0.6]{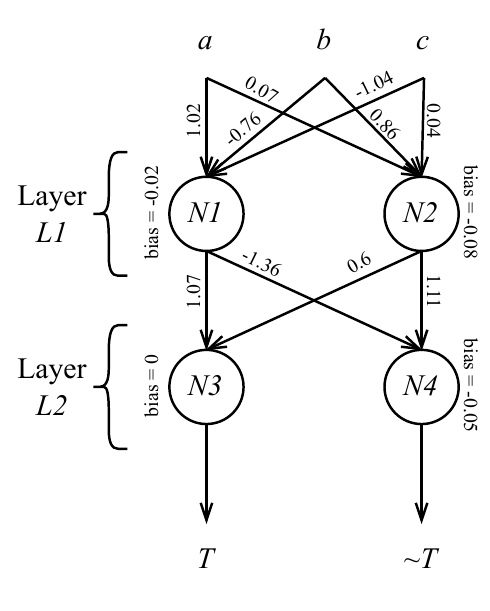}
    \vspace{-3mm}
    \caption{Model structure built and fitted in Listing~\ref{lst:first}}
    \label{fig:badDNN}
\end{wrapfigure}

We now describe how \tool worked, under the hood, to detect the bug \via \metallaxis' default formula.
Figure~\ref{fig:badDNN} depicts the structure of the model constructed and fitted in Listing~\ref{lst:first}.
Each edge is annotated with its corresponding weight and the nodes are annotated with their bias values.
The nodes are using ReLU as the activation function.
In this model, the output $T$ is intended to be greater than the other output if $a$, $b$, and $c$ form a triangle, and $
\sim$$T$ should be greater than or equal to the other output, otherwise.

Table~\ref{tab:motivating} shows an example of how \tool localizes the bug in the model depicted in Figure~\ref{fig:badDNN}.
In the first two columns, the table lists the two layers, and within each layer, the neurons.
For each neuron three mutators are applied, \ie, halving weight values, halving bias value, and replacing the activation function.
More mutators are implemented in \tool, but here, for the sake of simplicity, we only focus on 3 of them and also restrict ourselves to only one activation function replacement, \ie, ReLU vs. softmax.

As we saw in Courtney's example, she had a test dataset $T$ with 6 data points which initially resulted in 33\% accuracy.
These six data points are denoted T1, ..., T6 in Table~\ref{tab:motivating}, where correctly classified ones are colored green, whereas misclassified data points are colored rose.
\tool generates 12 mutants for the model of Figure~\ref{fig:badDNN}, namely, M1, ..., M12.
Each mutant is a variant of the original model.
For example, M1 is the same as the model depicted in Figure~\ref{fig:badDNN}, except the weights of the incoming edges to neuron N1 are halved, \ie, 0.51, -0.38, and -0.52 from left to right, while M9 is the same as the model depicted in Figure~\ref{fig:badDNN}, except that the activation functions for N3 and N4 are {\small \texttt{softmax}} instead of {\small \texttt{relu}}.
After generating the mutants, \tool applies each mutant on the inputs T1, ..., T6 and compares the results to that of the original model.
For each data point T1, ..., T6, if the result obtained from each mutant M1, ..., M12 is different from that of the original model, we put a bullet point in the corresponding cell.
For example, two bullets points in the row for M3 indicate that the mutant misclassifies the two data points that used to be correctly classified, while other data points, \ie, T1, ..., T4, are misclassified as before.
Next, \tool uses SBI formula~\cite{liblit2005scalable} to calculate suspiciousness values for each mutant $m\in\{$M1, ..., M12$\}$, individually.
These values are reported under the one but last column in Table~\ref{tab:motivating}.
Lastly, \tool takes the maximum of the suspiciousness values of the mutants corresponding to a layer and takes it as the suspiciousness value of that layer (c.f. Eq. \ref{eq:metallaxis:sbi} in \cref{sec:background}).
In this particular example, layer L1 gets a suspiciousness value of 0, while L2 gets a suspiciousness value of 1.
Thus, \tool ranks L2 before L1 for user inspection and for each layer it sorts the mutants in descending order of their suspiciousness values, so that the user will understand what change impacted most the originally correctly classified data points.
In this case, M12 and M9 wind up at the top of the list, and as we saw in Courtney's story, the information associated with the mutations helped fixing the bug.

\section{Proposed Approach}\label{sec:approach}
\begin{figure}[t]
    \centering
    \includegraphics[scale=0.5]{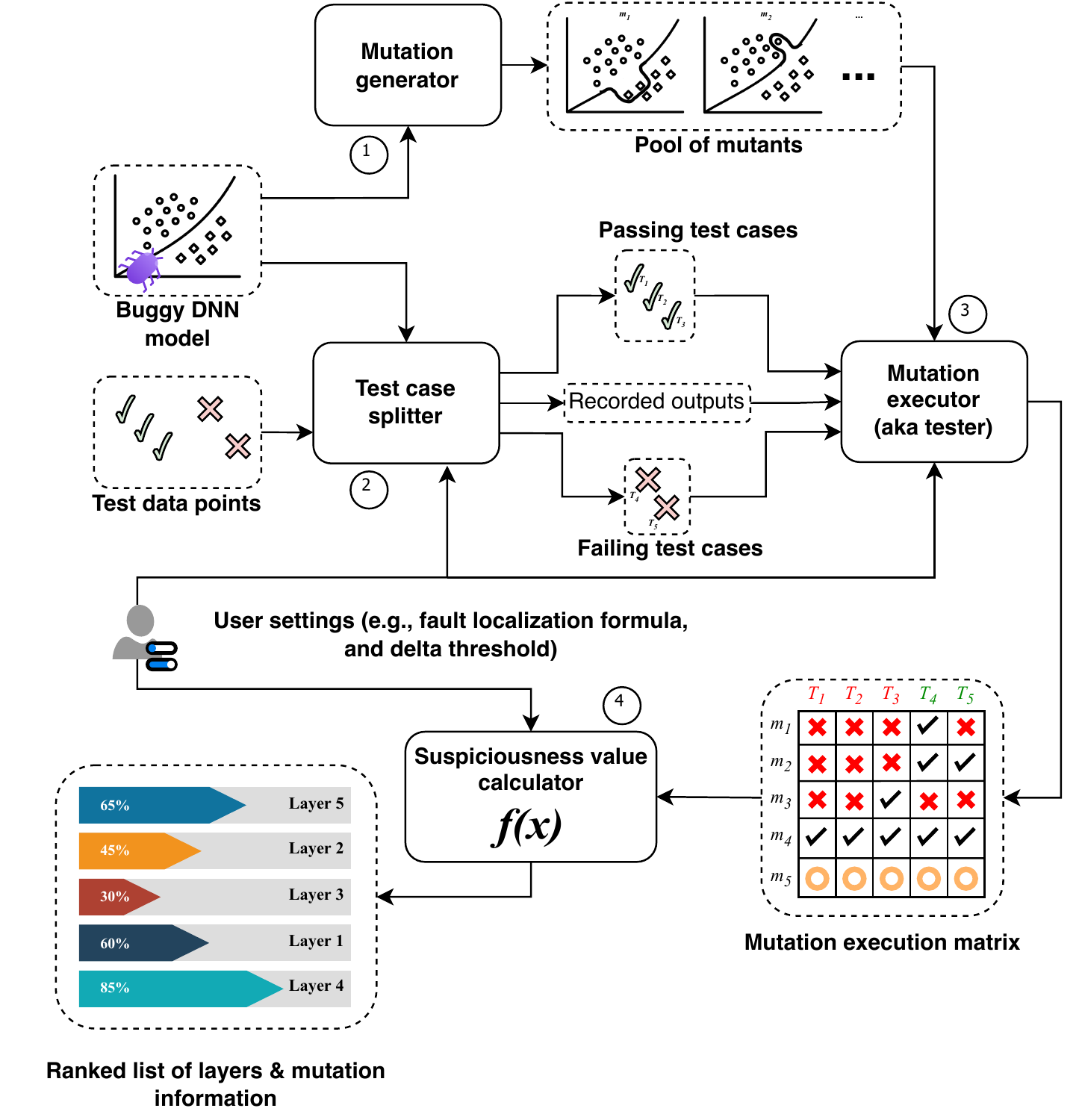}
    \caption{{\scriptsize \textsf{deepmufl}} architecture. Processes are denoted using solid round rectangles, while data and artifacts are represented as dotted round rectangles. Arrows represent flow of control and information.}
    \label{fig:arch}
\end{figure}
Our technique \tool comprises four components: (1) mutation generator, (2) test case splitter, (3) mutation executor/tester, and (4) suspiciousness calculator.
Figure~\ref{fig:arch} depicts these components as processes, numbered accordingly, taking inputs and producing outputs.
Mutation generator (marked \textcircled{\textsf{\scriptsize{1}}} in Figure~\ref{fig:arch}), applies \allMutators mutators on all the layers of the input buggy DNN, so as to generate a pool of mutants, \ie, variants of the original, buggy DNN model with small perturbations, \eg, replacing activation function of a layer.
Test case splitter (marked \textcircled{\textsf{\scriptsize{2}}} in the figure), applies the original buggy DNN on a given set of test data points, \ie, test cases (or input values) paired with their expected output values, so as to partition the set into two subset, namely \textit{passing test cases} and \textit{failing test cases}.
Passing test cases are referred to as those input values for which the expected output matches that of produced by the original model, whereas failing test cases are referred to as those input values for which the expected output does not match that of produced by the model.
This component also stores the output of the original model on each of the test cases.
Next, mutation executor (which is also called mutation tester, marked \textcircled{\textsf{\scriptsize{3}}} in the figure) applies the generated mutants on each of the passing and failing test cases and the results are compared to that of the original model recorded in the previous step.
This amounts to a mutation execution matrix that is used to calculate suspiciousness values for each layer in the model (marked \textcircled{\textsf{\scriptsize{4}}} in the figure).
The user may instruct \tool to use a specific fault localization formula, \eg, \muse or \metallaxis with SBI or Ochiai, for calculating suspiciousness values.
The layers are then ranked based on the calculated suspiciousness values for user inspection.
The ranked list is accompanied with the information about the mutations conducted on each layer to facilitate debugging.

\subsection{Mutation Generator}\label{sec:gen}
Mutation generator component receives the original, buggy DNN model and generates as many variants of the model, \ie, mutants, as possible, by systematically mutating every elements of the input model.
% \texttt{.h5} binary files contain all the information about model structure, learned parameters (\ie, weight, biases, and kernel matrices), and activation functions.
% The mutants are organized in a way that the set of all mutants that are generated by mutating a given model element $e$ can be queried efficiently.
This component implements \allMutators mutators.
Mutators can be viewed as transformation operators that when applied to a given element, \eg, a neuron or a layer, in the model, returns a new variants of the model with that particular element mutated.
Table~\ref{tab:mutators} lists all the mutators implemented in \tool, the types of model elements on which they can operate, and the way each mutator affects the target element.
These mutators are inspired by the ones implemented in the existing mutation analysis systems, \eg,~\cite{humbatova2021deepcrime,hu2019deepmutation++,ma2018deepmutation,coles2016pit,ammann2016introduction,just2011major,ma2005mujava,schuler2009javalanche}, to name a few.
Ma \etal~\cite{ma2018deepmutation}, and Hu \etal~\cite{hu2019deepmutation++}, define so-called \textit{model-level} mutators that also operate on pre-trained models.
Direct reuse of all of their mutators was not possible, as those mutators depend on random values which would introduce a source of non-determinism in \tool: mutating the same model element with random values, \eg, Gaussian fuzzing, as in~\cite{ma2018deepmutation}, would yield a different mutant each time, making \tool to produce different outputs on each run for the same model.
In general, as far as MBFL is concerned, using variable values (whether it is deteministic or not), instead of the current hard-coded ones, for mutation of weights would not bring about any benefit, as the goal here is to \textit{break} the model in some way and observe its impact on originally failing and passing test cases.

We argue that not all model bugs could be emulated using mutators at the level of pre-trained models, \eg, missing batch normalization, but the mutators listed in Table~\ref{tab:mutators} are sufficient for emulating a subset of such bugs, \eg, wrong activation function or missing/redundant layer.
Please see \cref{sec:scope} for a more detailed discussion on supported bugs.

Mutation generation in \tool is done directly on the trained model and there is no need for retraining the model.
This makes \tool quite fast and perhaps more attractive from a practical point of view.
However, this comes with a risk of not being traceable, \ie, a mutation on pre-trained {\small \texttt{.h5}} model does not directly correspond to a line of source code for the user to inspect.
In the Keras programs that we studied, this limitation was mitigated by the fact that the models with {\small \texttt{Sequential}} architecture were implemented using a well-understood structure and mapping layer numbers/identifiers in \tool's reports to source code was trivial.
In a future work, with the help of simple auto-generated annotations, \eg, for lexical scoping of the code snippet for model declaration, we will extend \tool to automatically map layer numbers/identifiers in its reports to actual lines of source code.

Humbatova \etal~\cite{humbatova2021deepcrime} argue about the importance of mutators emulating real DNN faults.
We acknowledge that mutators emulating real faults would help generating more informative reports that would also give hints on how to fix the program.
However, unlike mutation analysis, the main objective of an MBFL technique is to assign suspiciousness values to the program elements which can, in theory, be done using any kind of mutator, whether or not it makes sense from the standpoint of a developer.
It is worth noting that the alternative design decision of using DeepCrime~\cite{humbatova2021deepcrime} as a mutation generation engine for \tool would result in finding more bugs than the current version of \tool, \eg, bugs related to training hyper-parameters or training dataset, but such a design is expected to be impacted by the nondeterminacy inherent in training process and, given the fact that we do not employ any training data selection, would be significantly slower due to numerous re-training.
Nevertheless, finding more bugs would be an incentive for exploring this alternative as a future work.

\subsection{Test Case Splitter}
Before we introduce this component, we need to clarify certain terminology.
\begin{definition}
A data point in a testing dataset for a DNN model is defined to be a pair of the form $(I, O)$, where $I\in\mathbb{R}^m$ and $O\in\mathbb{R}^n$, with $m$ and $n$ being positive integers.
In this paper $I$ is called test case, test input, or input, while $O$ is called expected output or ground-truth value.
\end{definition}

Given a test dataset, test case splitter component applies the original model on each of the test cases for the data points and checks if the model output matches to the expected output.
If the two outputs match, then the corresponding test case will be marked as \textit{passing}, otherwise it will be marked as \textit{failing}.
This component also records the output produced by the original model to be used during impact analysis, described below.

\subsection{Mutation Executor (Mutation Tester)}\label{sec:exec}
We start describing this component with a definition.
\begin{definition}\label{def:impact}
A mutation execution matrix $\mathcal{E}$ is a $k\times l$ matrix, where $k$ is the number of generated mutants, while $l$ is the number of test cases.
Each element $\mathcal{E}_{i,j}$ in the matrix is a member of the set $\{\impacts,\noimpact,\nonviable\}$, wherein \impacts indicates that the $i^\mathrm{th}$ mutant impacts $j^\mathrm{th}$ test case, whereas \noimpact indicates that the mutant does not affect the test case.
\nonviable denotes a nonviable mutant, \ie, a mutant that fails when loading or applying it on a test case.
Such mutants might be generated, \eg, due to creating a shape error~\cite{tensorflow2016abadi} after the mutation.
\end{definition}
% It is worth noting that in most of the cases, if for some $i$ and $j$, $\mathcal{E}_{i,j}=\nonviable$, then for all $j'$ we will have $\mathcal{E}_{i,j'}=\nonviable$.
% In other words, a nonviable mutant will always fail in one way or another regardless of the test case that we feed to it.
\begin{table}
    \centering
    \caption{Summary of the \allMutators mutators implemented in \toolSmall}\label{tab:mutators}
    \resizebox{\columnwidth}{!}{
        \begin{tabular}{c||p{9.5cm}}
            \textbf{Mutator Class} & \multicolumn{1}{c}{\textbf{Description}} \\ \hline\hline
            \multirow{2}{*}{\mathWeight} & Add/subtract 1 to/from the weights of a given neuron and multiply/divide them to/by 2. Targets {\small \textsf{Dense}} and {\small \textsf{SimpleRNN}} layers.\\ \hline
            \multirow{3}{*}{\mathWeightConv} & Add/subtract 1 to/from the weights of a convolution layer and multiply/divide them to/by 2. Targets subclasses of {\small \textsf{Conv}}, \ie, {\small \textsf{Conv1D}}, {\small \textsf{Conv2D}}, \etc. \\ \hline
            \multirow{2}{*}{\mathActWeight} & Add/subtract 1 to/from the activation weights of a (rolled) recurrent layer and multiply/divide them to/by 2. Targets {\small \textsf{SimpleRNN}} layers.\\ \hline
            \multirow{2}{*}{\mathLSTMInWeight} & Add/subtract 1 to/from the input weights of an LSTM layer and multiply/divide them to/by 2. Targets {\small \textsf{LSTM}} layers.\\ \hline
            \multirow{2}{*}{\mathLSTMForgetWeight} & Add/subtract 1 to/from the forget weights of an LSTM layer and multiply/divide them to/by 2. Targets {\small \textsf{LSTM}} layers.\\ \hline
            \multirow{2}{*}{\mathLSTMCellWeight} & Add/subtract 1 to/from the cell weights of an LSTM layer and multiply/divide them to/by 2. Targets {\small \textsf{LSTM}} layers.\\ \hline
            \multirow{2}{*}{\mathLSTMOutWeight} & Add/subtract 1 to/from the output weights of an LSTM layer and multiply/divide them to/by 2. Targets {\small \textsf{LSTM}} layers.\\ \hline
            \multirow{2}{*}{\mathBias} & Add/subtract 1 to/from the bias value of neuron and multiply/divide them to/by 2. Targets {\small \textsf{Dense}} and {\small \textsf{SimpleRNN}} layers.\\ \hline
            \deleteLayer & Deletes a {\small \textsf{Dense}} layer.\\ \hline
            \duplicateLayer & Duplicates a {\small \textsf{Dense}} layer.\\ \hline
            \multirow{2}{*}{\mathConvBias} & Add/subtract 1 to/from the bias value of a convolution layer and multiply/divide them to/by 2. Target subclasses of {\small \textsf{Conv}}.\\ \hline
            \multirow{2}{*}{\mathLSTMInBias} & Add/subtract 1 to/from the input bias of an LSTM layer and multiply/divide them to/by 2. Target {\small \textsf{LSTM}} layers.\\ \hline
            \multirow{2}{*}{\mathLSTMForgetBias} & Add/subtract 1 to/from the forget bias of an LSTM layer and multiply/divide them to/by 2. Target {\small \textsf{LSTM}} layers.\\ \hline
            \multirow{2}{*}{\mathLSTMCellBias} & Add/subtract 1 to/from the cell bias of an LSTM layer and multiply/divide them to/by 2. Target {\small \textsf{LSTM}} layers.\\ \hline
            \multirow{2}{*}{\mathLSTMOutBias} & Add/subtract 1 to/from the output bias of an LSTM layer and multiply/divide them to/by 2. Target {\small \textsf{LSTM}} layers.\\ \hline
            \multirow{3}{*}{\actRep} & Replaces the activation of a neuron with a different one. Targets all layers with activation function in their configuration, \eg, {\small \textsf{Dense}}, {\small \textsf{Conv2D}}, \etc.\\ \hline
            \mathPoolSz & Increase/decrease pool size by 1, if applicable.\\ \hline
            \mathStrides & Increase/decrease strides by 1, if applicable.\\ \hline
            \mathKernelSz & Increase/decrease kernel size by 1, if applicable.\\ \hline
            \mathFilters & Increase/decrease filters by 1, if applicable.\\ \hline
            \multirow{2}{*}{\paddingRep} & Replace {\small \textsf{valid}} padding with {\small \textsf{same}}, \viceversa. Targets subclasses of {\small \textsf{Conv}}.\\ \hline
            \multirow{2}{*}{\recActRep} & Replace the recurrent activation of a layer. Targets {\small \textsf{SimpleRNN}} layers.\\ \hline
        \end{tabular}
    }
\end{table}

Mutation executor component construct mutation execution matrix by applying each of the generated mutants (see~\cref{sec:gen}) on the failing and passing test cases to determine which of the test cases are impacted by which mutants.
The impact of mutation on test cases is measured using two types of impacts, \ie, \textit{type 1} impact and \textit{type 2} impact, defined below.
\begin{definition}\label{def:types}
Given a DNN model $\mathcal{M}$, its mutated version $\mathcal{M}'$, and a data point $(I, O)$, we define the two types of impacts:
\begin{itemize}
    \item Type 1: Mutant $\mathcal{M}'$ impacts the test case $I$ if $\mathcal{M}(I)=O$ but  $\mathcal{M}'(I)\neq O$, or $\mathcal{M}(I)\neq O$ but $\mathcal{M}'(I)=O$. In other words, type 1 impact tracks pass to fail and fail to pass test cases, \ala \muse~\cite{moon2014ask}.
    \item Type 2: Mutant $\mathcal{M}'$ impacts the test case $I$ if $\mathcal{M}(I)\neq\mathcal{M}'(I)$.
\end{itemize}
In this definition, $\mathcal{M}(I)$ or $\mathcal{M}'(I)$ denotes the operation of applying model $\mathcal{M}$ or $\mathcal{M}'$ on the test case $I$.
\end{definition}

It is worth noting that checking for equality of two values can be tricky for regression models, as those models approximate the expected values.
To work around this problem, \tool compares values obtained from regression models using a user-defined delta threshold, \ie, the values are deemed equal if their absolute difference is no more than a threshold.
By default, \tool uses a threshold of 0.001.
This is the approach adopted by well-known testing frameworks for comparing floating-point values~\cite{bib:junit,bib:TestNG}.
Also, whether \tool uses type 1 or type 2 impact is a user preference and is determined alongside the threshold.

\subsection{Suspiciousness Value Calculator}
Armed with the above definitions, we can now give concrete definitions for the terms used in Eq.~\ref{eq:metallaxis:sbi},~\ref{eq:metallaxis:ochiai}, and~\ref{eq:muse}, specialized to DNNs.
\begin{itemize}
    \item Given a model element, \ie, neuron or a layer, $e$, $M(e)$ is defined to be the set of all mutants generated by mutating $e$. These sets are produced by mutation generator process.
    \item Assuming that $m$ is a mutant on the element $e$, $T_f(m,e)$ (or $T_p(m,e)$) is defined as the set of failing (or passing) test cases that are impacted in type 1 or type 2 fashion by $m$. More concretely, $T_p(m,e)=\{t\mid \mathcal{E}_{m,t}=\checkmark\wedge t~\mathrm{is~passing}\}$, similarly $T_f(m,e)=\{t\mid \mathcal{E}_{m,t}=\checkmark\wedge t~\mathrm{is~failing}\}$. These two sets are defined using a quite similar notation; the readers are encouraged to review Definitions~\ref{def:impact} and \ref{def:types} to avoid confusion.
    \item $T_f$ (or $T_p$) is the set of originally failing (or passing) test cases. These sets are constructed by test case splitter.
    \item $F\leadsto P$ (or $P\leadsto F$), for a model element $e$, is defined as the set of originally failing (or passing) test cases that turned into passing (or failing) as a result of some mutation on $e$. More concretely, assuming the execution matrix $\mathcal{E}$ is constructed using type 1 impact, $F\leadsto P$ is defined as $\{t\mid t~\mathrm{is~failing}\wedge \exists m\in M(e)\cdot \mathcal{E}_{m,t}=\checkmark\}$. Similarly, $P\leadsto F$ is defined as $\{t\mid t~\mathrm{is~passing}\wedge \exists m\in M(e)\cdot \mathcal{E}_{m,t}=\checkmark\}$.
    In other words, these sets track all the failing/passing test cases that are \textit{type 1 impacted} by some mutant of a given element. These two sets are defined using a quite similar notation; the readers are encouraged to review Definitions~\ref{def:impact} and \ref{def:types} to avoid confusion. 
\end{itemize}

Having specialized definitions for the terms used in the fault localization formulas described earlier, we are now able to calculate suspiciousness values for the elements in a DNN model.
Guided by the user preferences, \tool calculates all the values for $|T_f(m,e)|$, \etc., and plugs into the specified formula to calculate suspiciousness values for the elements.
It is worth noting that if all the mutants generated for a given element are nonviable, \muse formula (Eq.~\ref{eq:muse}) and all the variants of \metallaxis (\eg, Eq.~\ref{eq:metallaxis:sbi}), by definition, will return 0 as the suspiciousness value for the element.
Nonviable mutants do not contribute toward localizing the bug, therefore they are considered \textit{overhead} to the fault localization process.
Fortunately, based on our observations in our dataset of buggy DNNs, nonviable mutants are rare.

Equivalent mutants are another source of overhead for \tool.
Currently, we do not have any means of detecting equivalent mutants, but we argue that these mutants do not impact MBFL results, as they are equivalent to the original model and do not impact any passing or failing test cases.
%: only 1,150 out of 119,089 generated mutants happened to be nonviable.

\section{Supported DNN Bugs}\label{sec:scope}
Due to the complex nature of DNN bugs, and MBFL itself, we do not hope to give a formal account of what types of DNN bugs \tool is capable of localizing.
Instead, we attempt to provide as accurate description of the supported bugs as possible and discuss the way such bugs manifest in DNN programs.
The discussion given in this section leverages the characterization of DNN bugs provided by previous research~\cite{humbatova2020taxonomy,islam2019comprehensive,zhang2018empirical}.

As we mentioned earlier, current version of \tool operates on pre-trained Keras \texttt{Sequential} models.
This means that much of the information, such as training hyper-parameters and whether or not the input data is normalized, has already been stripped away from the input to \tool, and the current version of the technique is not capable of detecting any bug related to training process, \eg, training data and hyper-parameters.
Moreover, a pre-trained model does not contain bugs related to tensor shapes (as otherwise, the training would fail with shape errors), and since \tool does not receive the source code of the buggy model as input, bugs related to GPU usage and API misuse are also out of the reach of the technique, by definition.
This leaves us with the so-called \textit{model bugs}~\cite{humbatova2020taxonomy} the extent to which \tool is capable of localizing is explicated below.
The four model bug sub-categories are represented with identifiers SC1, ..., SC4 in the rest of this paper for ease of reference.
\begin{itemize}
    \item \textbf{SC1: Activation function}. These bugs are related to the use of wrong activation function in a layer. We observed that \tool detects this type of bugs and it also gives actionable, direct fixes.
    \item \textbf{SC2: Model type or properties}. These bugs include wrong weight initialization, wrong network architecture, wrong model for the task, \etc. Through altering the weights and biases in layers, \tool detects weight/bias initialization bugs and pinpoint the location of the bug, but the bug report produced by the tool does not provide helpful information for fixing.% the bug.
    \item \textbf{SC3: Layer properties}. These bugs include wrong filter/kernel/stride size, sub-optimal number of neurons in a layer, wrong input sample size, \etc. \tool detects and pinpoints the bugs related to filter/kernel/stride size and sub-optimal number of neurons. We observed that, the former case sometimes produce non-viable mutants. In the cases where \tool produced viable mutants, effective MBFL takes place and it has been able to pinpoint the bug location and provide explanation on how to fix it. In the latter case, \tool was able to pinpoint the bug location, but the bug report does not give helpful information on how to fix the bugs in this sub-category.
    \item \textbf{SC4: Missing/redundant/wrong layer}. These bugs include missing/extra one dense layer, missing dropout layer, missing normalization layer, \etc. By mutating the layers adjacent to the missing layer, or deleting the redundant layer, \tool detects and pinpoints the location of the missing/culprit layer, and in most of the cases, it provides useful information on how to fix such bugs.
\end{itemize}

By manually examining the bug descriptions provided by the programmers in our dataset of bugs, and also referring to the previous work on DNN bugs and root cause characterization~\cite{islam2019comprehensive}, these bugs might manifest as low test accuracy/MSE, constant validation accuracy/MSE/loss during training, NaN validation accuracy/MSE/loss during training, dead nodes, vanishing/exploding gradient, and saturated activation.

At this point, we would like to emphasize that \tool is not intended to repair a model, so if a mutation happens to be the fix for the buggy model, the model has to be retrained from scratch so that correct weights and biases will be calculated.

\section{Evaluation}\label{sec:evaluation}
We evaluate \tool and compare it to state-of-the-art static and dynamic DNN fault localization techniques, by investigating the following research questions (RQs).
\begin{itemize}
    \item \textbf{RQ1 (Effectiveness)}:
    \begin{enumerate}
        \item How does \tool compare to state-of-the-art tools in terms of the number of bugs detected?
        \item How many bugs does \tool detect from each sub-category of model bugs in our dataset and how does that compare to state-of-the-art tools?
        \item What are the overlap of detected bugs among \tool and other fault localization techniques?
    \end{enumerate}
    \item \textbf{RQ2 (Efficiency)}:
    \begin{enumerate}
        \item What is the impact of mutation selection on the effectiveness and efficiency of \tool?
        \item How does \tool compare to state-of-the-art tools in terms of end-to-end fault localization time?
    \end{enumerate}
    
    % \item \textbf{RQ3 (Debugging Effort Reduction)}: How does \tool compare to state-of-the-art in terms of the expected reduction of debugging effort?
\end{itemize}

\subsection{Dataset of DNN Bugs}\label{sec:evaluation:dataset}
To evaluate \tool and compare it to state-of-the-art DNN fault localization techniques, we queried StackOverflow Q\&A website for posts about Keras that had at least one accepted answer.
Details about the SQL query used to obtain the initial list of posts is available online~\cite{replication}.
The query resulted in \soPosts posts that we manually sieved through to find the programs with model bugs.
Specifically, we kept the bugs that satisfied the following conditions.
\begin{itemize}
    \item Implemented using {\small \texttt{Sequential}} API of Keras,
    \item The bug in the program was a \textit{model bug} supported by \tool as described in \cref{sec:scope}, and
    % \begin{itemize}
    %     \item Incorrect activation function,
    %     \item Sub-optimal number neurons in a layer,
    %     \item Missing/redundant/wrong layer, \eg, missing batch normalization or an extra dense layer,
    %     \item Incorrect layer properties, \eg, kernel size, \etc.
    %     \item Wrong initialization
    % \end{itemize}
    \item The bug either had training datatset available in the post in some form (\eg, hard-coded, clearly described in the body of the post, or a link to the actual data was provided) or we could see the described error using synthetic data obtained from {\small \texttt{scikit-learn}}'s dataset generation API.
\end{itemize}
This resulted in 102 bugs and we paired each bug with a fix obtained from the accepted answer to the question.
We further added 7 bugs from DeepLocalize dataset~\cite{wardat2021deeplocalize} that are also coming from StackOverflow and we paired these bugs also with their fixes that are obtained from the most up-voted answer.
Thus, we ended up with 109 bugs in total.
To the best of our knowledge, this is the largest dataset of model bugs obtained from StackOverflow and it overlaps with the existing DNN bug datasets from previous research~\cite{nikanjam2021automatic,morovati2022bugs}.
Our bug dataset contains 85 classifiers (45 fully-connected DNNs, 29 CNNs, and 11 RNNs) and 24 regression models (19 fully-connected DNNs, 3 CNNs, and 2 RNNs).
And each category has at least one example of model bugs.
Therefore, we believe that our dataset is greatly representative of model bugs, \ie, the bugs supported by \tool (and other tools that support this type of bugs), as we have examples of each sub-category of bug in various locations of the models for various regression and classification tasks.

After loading the training dataset for the bugs, we fitted the buggy models three times and stored them in {\small \texttt{.h5}} file format separately.
The repetition was conducted to take randomness in training into account.
Randomness in data generation was mitigated by using deterministic random seeds.
For fault localization purposes, we used the test dataset, and if it was not available, we used the training dataset itself.
When we had to use synthesized data points, we deterministically splitted the generated set of data into training and testing datasets.

% At this point, we have not included bugs from other sources, \eg, GitHub, as in our experience, the DNN bugs in GitHub are usually trained/tested using propriety datasets and we did not add the ones that are trained/tested using canonical datasets, such as MNIST~\cite{deng2012mnist}, as they are already represented by the StackOverflow bugs in our dataset.
% GitHub was not the only source where we could obtain DNN bugs, \eg, Zhang \etal~\cite{zhang2021autotrainer} use a large dataset of bugs for evaluating their repair tool.
% However, we could not reuse their bug dataset, because building on previous studies in software debugging, \eg, that of Monperrus~\cite{monperrus2014critical}, we strive to use a dataset of bugs that satisfies at least these three conditions: (1) be curated based on a transparent methodology; (2) have ground-truth patches, \ie, bug location be known; (3) represent real bugs, \ie, they are not seeded bugs.
% Unfortunately, we are not sure how the bugs from \autotrainer dataset are obtained and, more importantly, where the bugs are actually located.

\subsection{Baseline Approaches and Measures of Effectiveness}\label{sec:evaluation:baseline}
In RQ1 and RQ2, we compare five different configurations of \tool to recent static and dynamic DNN fault localization tools.
The five configurations of \tool are as follows.
\begin{itemize}
    \item \textbf{\metallaxis}~\cite{papadakis2015metallaxis}: In this setting, we use the \metallaxis formula to calculate suspiciousness values of model elements. \metallaxis, by default, uses SBI~\cite{liblit2005scalable} to calculate suspiciousness values for individual mutants. A recent study~\cite{li2017transforming} provides empirical evidence on the superiority of Ochiai~\cite{abreu2006evaluation} over SBI when used within \metallaxis formula. Thus, we considered the following four combinations: \textit{type 1 impact}: (1) SBI formula (\ie, Eq.~\ref{eq:metallaxis:sbi}); (2) Ochiai formula (\ie, Eq.~\ref{eq:metallaxis:ochiai}), and \textit{type 2 impact}: (3) SBI formula (\ie, Eq.~\ref{eq:metallaxis:sbi}); (4) Ochiai formula (\ie, Eq.~\ref{eq:metallaxis:ochiai}).
    \item \textbf{\muse}~\cite{moon2014ask}: We used the default formula of \muse to calculate the suspiciousness of model elements. For this, only type 1 impact is considered, as the heuristics behind \muse are defined based on type 1 impact.
\end{itemize}

Our technique follows a more traditional way of reporting root causes for the bugs~\cite{papadakis2015metallaxis,moon2014ask,li2017transforming,abreu2007accuracy,naish2011model,jones2002visualization,wong2016survey}, in that it reports a list of potential root causes ranked based on the likelihood of being responsible for the bug.
This allows the users find the bugs faster and spend less time reading through the fault localization report, which in turn increases practicality of the technique~\cite{parnin2011automated}.
We have used top-$N$, with $N=1$, metric to measure the effectiveness of \tool in RQ1 and RQ2.
Specifically, if the numbers of any of the buggy layers of the bug appeared in the first place in the output of \tool, we reported it as \textit{detected}, otherwise we marked the bug as \textit{not-detected}.
We emphasize that top-1 metric gives a strong evidence on the effectiveness of \tool, as the developers usually only inspect top-ranked elements, \eg, over 70\% of the developers only check top-5 ranked elements~\cite{kochhar2016practitioners}.
% In RQ3, we need actual rank of the bugs, so we used the rank generated by the \tool and other studied systems for calculating expected reduction in debugging efforts.

Our selection criteria for the studied fault localization techniques are: (1) availability; (2) reproducibility of the results reported in the original papers, so as to have a level of confidence on the correctness of the results reported here; and (3) support for \textit{model bugs} in our dataset, so that we can make a meaningful comparison to \tool.
Below we give a brief description of each of the selected tools, why we believe they support model bugs, and how we have interpreted their outputs in our experiments, \ie, when we regard a bug being detected by the tool.
%(in RQ1 and RQ2) and how we calculate \textit{worst-case rank}~\cite{kochhar2016practitioners,li2017transforming,zhang2017boosting} for the detected bugs.
% Table generated by Excel2LaTeX from sheet 'Config-class'
\begin{table}[t]
  \centering
  \caption{Effectiveness of different \toolSmall configurations and four other tools in detecting bugs from four sub-categories of model bugs}
  \resizebox{0.9\columnwidth}{!}{
    \begin{tabular}{c||r|r|r|r||r}
    \multicolumn{1}{c||}{\textbf{\tool configuration / tool}} & \multicolumn{1}{c|}{\textbf{SC 1}} & \multicolumn{1}{c|}{\textbf{SC 2}} & \multicolumn{1}{c|}{\textbf{SC 3}} & \multicolumn{1}{c||}{\textbf{SC 4}} & \multicolumn{1}{c}{\textbf{Total (detected)}} \\\hline\hline
    % & & & & & \\
    \textbf{\metallaxis SBI + Type 1} & 31    & 2     & 6     & \cellcolor{GrayTwo}3     & 42 \\\hline
    \textbf{\metallaxis Ochiai + Type 1} & 36    & 2     & \cellcolor{GrayTwo}7     & 2     & 47 \\\hline
    \textbf{\metallaxis SBI + Type 2} & 18    & 2     & 4     & 2     & 26 \\\hline
    \textbf{\metallaxis Ochiai + Type 2} & 29    & 2     & 4     & 2     & 37 \\\hline
    \textbf{\muse} & \cellcolor{GrayTwo}41    & \cellcolor{GrayTwo}3     & 6     & \cellcolor{GrayTwo}3     & \cellcolor{GrayTwo}53 \\\hline\hline
    \textbf{\neuralint} & 15 & 1 & 4 & 1 & 21 \\\hline
    \textbf{\dl} & 21 & 0 & 4 & 1 & 26 \\\hline
    \textbf{\dd} & 22 & 2 & 5 & 1 & 30 \\\hline
    \textbf{\umlaut} & 18 & 1 & 6 & 0 & 25 \\\hline\hline
    \textbf{Total (entire dataset)} & 80 & 4	& 17 &	8
    \end{tabular}%
    }
  \label{tab:deepmuflConfigClass}%
\end{table}%

\subsubsection{\neuralint}
A static fault localization tool that uses 23 rules to detect faults and design inefficiencies in the model.
Each rule is associated with a set of rules of thumb to fix the bug that are shown to the user in case the precondition for any of the rules are satisfied.
The five rules described in Section 4.2.1 of the paper target model bugs.
\neuralint produces outputs of the form $[\mathtt{Layer}~L~\mathtt{==>}~MSG]^*$, where $L$ is the suspicious layer number, and $MSG$ is a description of the detected issue and/or suggestion on how to fix the problem.
A bug is deemed \textit{detected} by this tool if it is located in the layer mentioned in the output message or the messages describe any of the root causes of the bug.
% We give a rank top-$M$ for a bug if it is detected by \neuralint and the tool has produced an output message with $M$ lines, and top-$N$, where $N$ is the number of layers in the buggy model under analysis, if the tool fails to detect the bug.

\subsubsection{\dl}
A dynamic fault localization technique that detects numerical errors during model training.
One of three rules described in Section III.D of the paper checks model bugs related to wrong activation function.
\dl produces a single message of the form $\mathtt{Batch}~B~\mathtt{Layer}~L:~MSG$, where $B$ is the batch number wherein the symptom is detected and $L$ and $MSG$ are the same as we described for \neuralint.
A bug is deemed \textit{detected} if it is located in the layer mentioned in the output message or the message describes any of the root causes of the bug.
% If a bug is detected by the tool, we regard it to appear in top-1 position, otherwise we regard it to appear in top-$N$ position, where $N$ is the number of layers in the model.

\subsubsection{\dd}
A tool similar to \dl, but with more bug pattern rules and a decision procedure to give actionable fix suggestions to the users based on the observations.
All 8 rules in Table 2 of the paper monitor the symptoms of model bugs.
Similar to \dl, \dd produces a single message of the form $\mathtt{Batch}~B~\mathtt{Layer}~L:~{MSG}_1~[\mathtt{OR}~{MSG}_2]$, where $B$ and $L$ are the same as described in \dl and ${MSG}_1$ and ${MSG}_2$ are two alternative solutions that the tool might suggest to fix the detected problem.
A bug is deemed \textit{detected} if it is located in the layer mentioned in the output message or the message describes any of the root causes of the bug.
% We regard a bug to appear in top-1 position if its successful detection is reported by a message of the form $\mathtt{Batch}~B~\mathtt{Layer}~L:~{MSG}_1$, top-2 if its successful detection is reported by a message of the form $\mathtt{Batch}~B~\mathtt{Layer}~L:~{MSG}_1~\mathtt{OR}~{MSG}_2$, and top-$N$ position, where $N$ is the number of layers in the model, otherwise.

\subsubsection{\umlaut}
A hybrid, \ie, a combination of static and dynamic, technique that works by applying heuristic static checks on, and injecting dynamic checks in, the program, parameters, model structure, and model behavior.
Violated checks raise error flags which are propagated to a web-based interface that uses visualizations, tutorial explanations, and code snippets to help users find and fix detected errors in their code.
All three rules described in Section 5.2 of the paper target model bugs.
The tool generates outputs of the form $[<{MSG_1}>\dots<{MSG_m}>]^*$, where $m>0$ and ${MSG}_i$ is a description of the problem detected by the tool.
A bug is deemed \textit{detected} if any of the messages match the fix prescribed by the ground-truth.
\subsection{Results}\label{sec:evaluation:results}
To answer RQ1, we ran \tool (using its five configurations) and four other tools on the 109 bugs in our benchmark.
We refer the reader to the repository~\cite{replication} for the raw data about which bug is detected by which tool, and here we describe the summaries and provide insights.

At top-1, \tool detects 42, 47, 26, 37, and 53 bugs using its \metallaxis SBI + Type 1, \metallaxis Ochiai + Type 1, \metallaxis SBI + Type 2, \metallaxis Ochiai + Type 2, and \muse, respectively, configurations.
Meanwhile \neuralint, \dl, \dd, and \umlaut detect 21, 26, 30, and 25, respectively, bugs.
Therefore, as far as the number of bugs detected by each technique is concerned, \muse configuration of \tool is the most effective configuration of \tool, significantly outperforming studied techniques, and \metallaxis Ochiai + Type 2 is the least effective one, outperformed by \dd.
An empirical study~\cite{li2017transforming}, which uses a specific dataset of traditional buggy programs, concludes that \metallaxis Ochiai + Type 2 is the most effective configuration for MBFL.
Meanwhile, our results for DNNs corroborates the theoretical results by Shin and Bae~\cite{shin2016theoretical}, \ie, we provide empirical evidence that in the context of DNNs \muse is the most effective MBFL approach.

Table~\ref{tab:deepmuflConfigClass} reports more details and insights on the numbers discussed above.
Specifically, it reports the number of bugs detected by each configuration of \tool an four other studied tools from each sub-category of model bugs present in our dataset of bugs.
As we can see from the upper half of the table, \muse is most effective in detecting bugs related to activation function (SC1), bugs related to model type/properties (SC2), and wrong/redundant/missing layer (SC4), while \metallaxis Ochiai + Type 1 configuration outperforms other configurations in detecting bugs related to layer properties (SC3).
Similarly, from bottom half of the table, we can see that other tools are also quite effective in detecting bugs related to activation function, with \dd being the most effective one among others.
We can also observe that \umlaut has been the most effective tool in detecting bugs related to layer properties.
As we can see, \muse configuration of \tool is consistently more effective than other tools across all bug sub-categories.

% Table generated by Excel2LaTeX from sheet 'tool-vs-tool'
\begin{table}[t]
  \centering
  \caption{The fraction of bugs detected by each \toolSmall configuration that are also detected by the other four tools}
  \resizebox{\columnwidth}{!}{
    \begin{tabular}{c||r|r|r|r||r}
          & \multicolumn{1}{c|}{\textbf{\neuralint}} & \multicolumn{1}{c|}{\textbf{\dl}} & \multicolumn{1}{c|}{\textbf{\dd}} & \multicolumn{1}{c||}{\textbf{\umlaut}} & \multicolumn{1}{c}{\textbf{Combined}} \\\hline\hline
    \textbf{\metallaxis SBI} & \multirow{2}{*}{23.81\% (10)} & \multirow{2}{*}{19.05\% (8)} & \multirow{2}{*}{21.43\% (9)} & \multirow{2}{*}{19.05\% (8)} & \multirow{2}{*}{59.52\% (25)} \\
    \textbf{+ Type 1} & & & & & \\\hline
    \textbf{\metallaxis Ochiai} & \multirow{2}{*}{27.66\% (13)} & \multirow{2}{*}{25.53\% (12)} & \multirow{2}{*}{21.28\% (10)} & \multirow{2}{*}{17.02\% (8)} & \multirow{2}{*}{57.45\% (27)} \\
    \textbf{+ Type 1} & & & & & \\\hline
    \textbf{\metallaxis SBI} & \multirow{2}{*}{15.38\% (4)} & \multirow{2}{*}{15.38\% (4)} & \multirow{2}{*}{11.54\% (3)} & \multirow{2}{*}{15.38\% (4)} & \multirow{2}{*}{46.15\% (12)} \\
    \textbf{+ Type 2} & & & & & \\\hline
    \textbf{\metallaxis Ochiai} & \multirow{2}{*}{13.51\% (5)} & \multirow{2}{*}{18.92\% (7)} & \multirow{2}{*}{21.62\% (8)} & \multirow{2}{*}{16.22\% (6)} & \multirow{2}{*}{48.65\% (18)} \\
    \textbf{+ Type 2} & & & & & \\\hline
    \textbf{\muse} & 22.64\% (12) & 22.64\% (12) & 28.3\% (15) & 24.53\% (13) & \cellcolor{GrayTwo}60.38\% (32) \\
    \end{tabular}%
    }
  \label{tab:toolVsTool}%
\end{table}%

Table ~\ref{tab:toolVsTool} provides further insights on the overlap of bugs detected by each variant of \tool and those detected by the other four tools.
Each value in row $r$ and column $c$ of this table, where $2\leq r\leq 5$ and $2\leq c\leq 6$, denotes the percentage of bugs detected by the \tool variant corresponding to row $r$ and tool corresponding to column $c$.
The values inside the parenthesis are the actual number of bugs.
For example, 8 out of 42, \ie, 19.05\%, of the bugs detected by \metallaxis SBI + Type 1 configuration of \tool are \textit{also} detected by \dl.
The last column of the table reports same statistics, except for all four of the studied tools combined. 
As we can see from the table, 60.38\% of the bugs detected by \muse configuration of \tool are already detected by one of the four tools, yet it detects 21 (=53-32) bugs that are not detected by any other tools.
This is because \tool approaches fault localization problem from a fundamentally different aspect giving it more flexibility.
Specifically, instead of looking for conditions that trigger a set of hard-coded rules, indicating bug patterns, \tool breaks the model using a set of mutators to observe how different mutation impact the model behavior.
Then by leveraging the heuristics underlying traditional MBFL techniques, it performs fault localization using the observed impacts on the model behavior.
Listing~\ref{lst:hardBugSC1} shows an example of a model bug that only \tool can detect.

\begin{center}
\begin{adjustbox}{width=0.8\columnwidth}
\lstset{language=python}
 \begin{lstlisting}[language = python,label=lst:hardBugSC1, basicstyle=\fontsize{6}{9}\selectfont, caption=Bug 48251943 in our dataset]
# load and split the dataset
# ...
model = Sequential()
model.add(Dense(4, input_dim=2, activation='relu'))
model.add(Dense(1, activation='relu'))
model.compile(loss='mean_squared_error', optimizer='sgd', metrics=['MSE'])
model.fit(X, Y, epochs=500, batch_size=10)
\end{lstlisting}
\end{adjustbox}
\end{center}

The problem with this regression model is that it does not output negative values, and given the fact that the dataset contains negative target values, the model achieves high MSE.
Since this model does not result in any numerical errors during training, \dl does not issue any warning messages, and since MSE decreases and the model does not show any sign of erratic behavior \dd does not detect the bug.
\umlaut's messages instruct adding softmax layer and checking validation accuracy which is clearly not related to the problem, because the bug is fixed by changing the activation function of the last layer to {\small \texttt{tanh}} and normalizing the output values.
Lastly, \neuralint issues an error message regarding incorrect loss function which also seems to be a false positive.

To answer RQ2, we ran \tool and the other four tools on a Dell workstation with Intel(R) Xeon(R) Gold 6138 CPU at 2.00 GHz, 330 GB RAM, 128 GB RAM disk, and Ubuntu 18.04.1 LTS and measured the time needed for model training as well as the MBFL process to complete.
We repeated this process four times, and in each round of \tool's execution, we randomly selected 100\% (\ie, no selection), 75\%, 50\%, and 25\% of the generated mutants for testing.
Random mutation selection is a common method for reducing the overhead of mutation analysis~\cite{pizzoleto2019systematic,wong1995reducing}.
During random selection, we made sure that each layer receives at least one mutants, so that we do not mask any bug.
The last row in Table~\ref{tab:mutationSelection} reports the average timing (of 3 runs) of MBFL in each round of mutation selection.
The table also reports the impact of mutation selection on the number of bugs detected by each configuration of \tool.
As we can see, in \muse configuration of \tool, by using 50\% of the mutants, one can halve the execution time and still detect 92.45\% of the previously detected bugs.
Therefore, mutation selection can be used as an effective way for curtailing MBFL time in DNNs.

For a fair comparison of \tool to state-of-the-art fault localization tools in terms of efficiency, we need to take into account the fact that \tool requires a pre-trained model as its input.
Thus, as far as the end-to-end fault localization time from an end-user's perspective is concerned, we want to take into consideration the time needed to train the input model in addition the time needed to run \tool.
With training time taken into account, \tool takes, on average, 1492.48, 1714.63, 1958.35, and 2192.4 seconds when we select 25\%, 50\%, 75\%, and 100\% of the generated mutants, respectively.
We also emphasize that the time for \dl and \dd varied based on whether or not they found the bug.
Given the fact that a user could terminate the fault localization process after a few epochs when they lose hope in finding bugs with these two tools, we report two average measurements for \dl and \dd: (1) average time irrespective of the fact that the tools succeed in finding the bug; (2) average time if the tools successfully finds the bug.
Unlike these two tools, the time for \neuralint and \umlaut does not change based on the fact that they detect a bug or not.
\dl takes on average 1244.09 seconds and it takes on average 57.29 seconds when the tool successfully finds the bug.
These numbers for \dd are 1510.71 and 11.05 seconds, respectively.
Meanwhile, \neuralint and \umlaut take on average 2.87 seconds and 1302.61 seconds to perform fault localization.
\subsection{Discussion}
% Table generated by Excel2LaTeX from sheet 'mutation selection'
% \begin{wraptable}{l}{0.6\columnwidth}
\begin{table}[t]
    \centering
  \caption{The impact of mutation selection on the effectiveness and execution time of \toolSmall}
  \resizebox{0.75\columnwidth}{!}{
    \begin{tabular}{c||r|r|r|r}
          & \multicolumn{4}{c}{\textbf{Selected mutants}} \\\cline{2-5}
          & \multicolumn{1}{c|}{\textbf{25\%}} & \multicolumn{1}{c|}{\textbf{50\%}} & \multicolumn{1}{c|}{\textbf{75\%}} & \multicolumn{1}{c}{\textbf{100\%}} \\\hline\hline
    \textbf{\metallaxis SBI + Type 1} & 37    & 41    & 42    & 42 \\\hline
    \textbf{\metallaxis Ochiai + Type 1} & 40    & 46    & 47    & 47 \\\hline
    \textbf{\metallaxis SBI + Type 2} & 25    & 26    & 26    & 26 \\\hline
    \textbf{\metallaxis Ochiai + Type 2} & 34    & 37    & 37    & 37 \\\hline
    \textbf{\muse} & 42    & 49    & 51    & 53 \\\hline
    \textbf{Time (s)} & 340.58 & 562.72 & 806.45 & 1,040.49 \\
    \end{tabular}%
}
  \label{tab:mutationSelection}%
\end{table}

It is important to note that while \tool outperforms state-of-the-art techniques in terms of the number of bugs detected in our dataset, it is not meant to replace them.
Our dataset only covers a specific type of bugs, \ie, model bugs, while other studied techniques push the envelope by detecting bugs related to factors like learning rate and training data normalization, which are currently outside of \tool's reach.
We observed that combining all the techniques results in detecting 87 of the bugs in our dataset; exploring ways to combine various fault localization approaches by picking the right tool based on the characteristics of the bug is an interesting topic for future research.
Moreover, depending on the applications and resource constraints, a user might prefer one tool over another.
For example, although \neuralint might be limited by its static nature, \eg, it might not be able analyze models that use complex computed values and objects in their construction, it takes only few seconds for the tool to conduct fault localization.
Thus, in some applications, \eg, online integration with IDEs, approaches like that of \neuralint might be the best choice.

A major source of overhead in an MBFL technique is related to the sheer number of mutants that the technique generates and tests~\cite{zhang2015scalability,pizzoleto2019systematic}.
Sufficient mutator selection~\cite{offutt1996experimental} is referred to the process of selecting a subset of mutators that achieve the same (or similar) effect, \ie, same or similar mutation score and same or similar number of detected bugs, but with smaller number of mutants generated and tested.
For the mutators of Table~\ref{tab:mutators}, so far, we have not conducted any analysis on which mutators might be redundant, as a reliable mutator selection requires a larger dataset that we currently lack.
We postpone this study as a future work.

Combining fault localization tools can be conducted with the goal of improving efficiency.
We see the opportunity in building faster, yet more effective, fault localization tools by predicting the likely right tool upfront for a given model or running tools one by one and moving on to the next tool if we have a level of confidence that the tool will not find the bug.
We postpone this study for a future work.

Lastly, we would like to emphasize that comparisons to the above-mentioned techniques in a dataset of bugs that \tool supports is fair, as the other tools are also designed to detect bugs in the category of model bugs.
However, making these tools to perform better than this, would require augmenting their current rule-base with numerous new rules, yet adding new rules comes with the obligation of justifying the generality and rationale behind them, which might be a quite difficult undertaking.
\tool, on the other hand, approaches the fault localization problem differently, allowing for more flexibility without the need for hard-coded rules.

\section{Threats to Validity}\label{sec:threats}
As with most empirical evaluations, we do not have a working definition of representative sample of DNN bugs, but we made efforts to ensure that the bugs we used in the evaluation is as representative as possible by making sure that our dataset has diverse examples of bugs from each sub-category of model bugs.

Many of the bugs obtained from StackOverflow did not come with accompanying training datasets.
To address this issue, we utilized the dataset generation API provided by {\small \texttt{scikit-learn}}~\cite{sklearn} to generate synthetic datasets for regression or classification tasks.
We ensured that the errors described in each StackOverflow post would manifest when using the synthesized data points and that applying the fix suggested in the accepted response post would eliminate the bug.
However, it is possible that this change to the training process may introduce new unknown bugs.
To mitigate this risk, we have made our bug benchmark publicly available~\cite{replication}.

Another potential threat to the validity of our results is the possibility of bugs in the construction of \tool itself, which could lead to incorrect bug localization.
To mitigate this, we make the source code of \tool publicly available for other researchers to review and validate the tool.

Another threat to the validity of our results is the potential impact of external factors, such as the stochastic nature of the training process and the synthesized training/testing datasets, as well as system load, on our measurements.
To address this, besides using deterministic seeds for dataset generation and splitting, we repeated our experiments with \tool three times.
Similarly, we also ran other dynamic tools three times to ensure that their results were not affected by randomness during training.
We did not observe any differences in effectiveness between the rounds for either \tool or the other studied techniques.
Additionally, we repeated the time measurements for each round, and reported the average timing, to ensure that our time measurements were not affected by system load.
Furthermore, judging whether or not any of the tools detect a bug requires manual analysis of textual description of the bugs and matching it to the tools; output messages which might be subject to bias.
To mitigate this bias, we have made the output messages by the tools available for other researchers~\cite{replication}.

Lastly, \tool uses a threshold parameter to compare floating-point values (see \cref{sec:exec}).
In our experiments, we used the default value of 0.001 and ensured that smaller threshold values yield the same results.

% \vspace{-3mm}
\section{Related Work}\label{sec:related}
\neuralint~\cite{nikanjam2021automatic} uses \textit{graph transformations}~\cite{heckel2006graph} to abstract away unnecessary details in the model and check the bug patterns directly on the graph.
While \neuralint is orders of magnitude faster than \tool, it proved to be less effective than \tool in our dataset.

\dl~\cite{wardat2021deeplocalize} and \dd~\cite{wardat2021deepdiagnosis} intercept the training process looking for known bug patterns such as numerical errors.
\dd pushes the envelope by implementing a decision tree that gives actionable fix suggestions based on the detected symptoms.
A closely related technique, \umlaut~\cite{schoop2021umlaut}, works by applying heuristic static checks on, and injecting dynamic checks in, various parts of the DNN program.
% This hybrid nature of \umlaut makes it a bit faster than \dl and \dd, in that some of the dynamic checks are off-loaded and are applied statically.
\tool outperforms \dl, \dd, and \umlaut in terms of the number of bugs detected.

\deepfd~\cite{cao2022deepfd} is a recent learning-based fault localization technique which frames the fault localization as a learning problem.
% Additionally, we observed that \dl times out a 5-hour time limit, in some cases, while \tool terminates within \toolAvgTime seconds, on average.
MODE~\cite{ma2018mode} and DeepFault~\cite{eniser2019deepfault} implement white-box DNN testing technique which utilizes suspiciousness values obtained \via an implementation of spectrum-based fault localization to increase the hit spectrum of neurons and identify suspicious neurons whose weights have not been calibrated correctly and thus are considered responsible for inadequate DNN performance.
% The tool also uses a suspiciousness-guided algorithm to synthesize new inputs, from correctly classified inputs, that increase the activation values of suspicious neurons.
MODE was not publicly available, but DeepFault was, but unfortunately it was hard-coded to the examples shipped with its replication package, so we could not make the tool work without making substantial modifications to it, not to mention that these techniques work best on ReLU-based networks and applying them on most of the bugs in our dataset would not make much sense.
% and we observed that we had to make substantial changes to the program to make it run on each of our benchmark programs.
% As a future work, we are planning to perform an extensive study to compare \tool to DeepFault.

Other related works are as follows.
PAFL~\cite{ishimoto2023pafl} operates on RNN models by converting such models into probabilistic finite automata (PFAs) and localize faulty sequences of state transitions on PFAs.
Sun \etal~\cite{sun2020explaining} propose DeepCover, which uses a variant of spectrum-based fault localization for DNN explainability.

\section{Conclusion}\label{sec:conclusion}
This paper revisits mutation-based fault localization in the context of DNN and presents a novel DNN fault localization technique, named \tool.
The technique is based on the idea of mutating a pre-trained DNN model and calculating suspiciousness values according to \metallaxis and \muse approaches, Ochiai and SBI formulas, and two types of impacts of mutations on the results of test data points.
\tool is compared to state-of-the-art static and dynamic fault localization systems~\cite{wardat2021deeplocalize,wardat2021deepdiagnosis,schoop2021umlaut,nikanjam2021automatic} on a benchmark of \bugsCount model bugs.
In this benchmark, while \tool is slower than the other tools, it proved to be almost two times more effective than them in terms of the total number of bugs detected and it detects 21 bugs that none of the studied tools were able to detect.
% Meanwhile, since \tool needs a pre-trained model as its input, it is slower than other tools.
%Additionally, our technique is at least 3 times more effective than other studied techniques in terms of the expected reduction in user debugging efforts.
We further studied the impact of mutation selection on fault localization time.
We observed that we can halve the time taken to perform fault localization by \tool, while losing only 7.55\% of the previously detected bugs.

% As a future work, we will investigate the impact of the quality of the training and test datasets, as well as the initial accuracy of the buggy model, on the effectiveness of \tool.
% We will also conduct an ablation study on the effectiveness of individual mutators and perform mutation selection in a heuristically guided manner rather than entirely random.
% Lastly, we are planning to employ regression test selection ideas from software testing to reduce the size of training dataset.
% In this way, we can efficiently use source code level MBFL to extend the scope of \tool and also mitigate current problems regarding traceability.

\section*{Acknowledgments}
The authors thank Anonymous ASE 2023 Reviewers for their valuable feedback.
We also thank Mohammad Wardat for his instructions on querying StackOverflow.
This material is based upon work supported by the National Science Foundation (NSF) under the grant \#2127309 to the Computing Research Association for the CIFellows Project.
This work is also partially supported by the NSF grants \#2223812, \#2120448, and \#1934884.
Any opinions, findings, and conclusions or recommendations expressed in this material are those of the authors and do not necessarily reflect the views of the NSF.

\bibliographystyle{IEEETran}
\bibliography{bibdb}

\end{document}